\documentclass[letterpaper]{article} % DO NOT CHANGE THIS
\usepackage{aaai25}  % DO NOT CHANGE THIS
\usepackage{times}  % DO NOT CHANGE THIS
\usepackage{helvet}  % DO NOT CHANGE THIS
\usepackage{courier}  % DO NOT CHANGE THIS
\usepackage[hyphens]{url}  % DO NOT CHANGE THIS
\usepackage{graphicx} % DO NOT CHANGE THIS
\usepackage{natbib}  % DO NOT CHANGE THIS AND DO NOT ADD ANY OPTIONS TO IT
\usepackage{caption} % DO NOT CHANGE THIS AND DO NOT ADD ANY OPTIONS TO IT
\usepackage{algorithm}
\usepackage{algorithmic}

\usepackage{tabularx}
\usepackage{booktabs}
\usepackage{newfloat}
\usepackage{listings}
\DeclareCaptionStyle{ruled}{labelfont=normalfont,labelsep=colon,strut=off} % DO NOT CHANGE THIS
\floatstyle{ruled}
\newfloat{listing}{tb}{lst}{}
\floatname{listing}{Listing}
\title{Decentralising LLM Alignment: A Case for Context, Pluralism, and Participation}
\author{
    Oriane Peter, Kate Devlin
}
\affiliations{
    Department of Digital Humanities, King’s College London, United Kingdom\\
    London WC2R 2LS, UK\\
    oriane.peter@kcl.ac.uk
}

\begin{document}

\maketitle

\begin{abstract}

Large Language Models (LLMs) alignment methods have been credited with the commercial success of products like ChatGPT, given their role in steering LLMs towards user-friendly outputs. However, current alignment techniques predominantly mirror the normative preferences of a narrow reference group, effectively imposing their values on a wide user base. Drawing on theories of the power/knowledge nexus, this work argues that current alignment practices centralise control over knowledge production and governance within already influential institutions. To counter this, we propose decentralising alignment through three characteristics: context, pluralism, and participation. Furthermore, this paper demonstrates the critical importance of delineating the context-of-use when shaping alignment practices by grounding each of these features in concrete use cases. 
This work makes the following contributions:
(1) highlighting the role of context, pluralism, and participation in decentralising alignment;
(2) providing concrete examples to illustrate these strategies; and
(3) demonstrating the nuanced requirements associated with applying alignment across different contexts of use.
Ultimately, this paper positions LLM alignment as a potential site of resistance against epistemic injustice and the erosion of democratic processes, while acknowledging that these strategies alone cannot substitute for broader societal changes.

\end{abstract}

\section{Introduction}
Self-reportedly, the \textit{``secret sauce"} behind OpenAI's commercial success was their introduction of a novel alignment algorithm for Large Language Models (LLMs) \cite{ouyangTrainingLanguageModels2022, heavenStoryHowChatGPT2023}. This algorithm enables developers to steer LLMs' outputs away from mere pre-trained text completion and towards a more conversational, user-friendly style. It allows developers to avoid, to some extent, the pitfalls faced by early LLMs, such as the replication of `toxic' views present in the model training data (e.g. \citealt{victorMicrosoftCreatedTwitter2016}). The success of OpenAI's Reinforcement Learning from Human Feedback (RLHF) method has catalysed the development of the LLM alignment research field. However, despite its increasing prevalence in research, the term `alignment' lacks a consistent definition, leading to critiques that it serves as a rhetorical placeholder more than as a precise concept \cite{kirkEmptySignifierProblem2023}. Such vagueness leaves significant ambiguity about \textit{what} models should be aligned \textit{to} and, critically, \textit{with whom} they should be aligned \cite{gabrielArtificialIntelligenceValues2020}.

Despite their intended global and diverse applications, LLMs predominantly generate outputs that are reflective of a narrow societal segment: Western, Educated, Industrialised, Rich, and Democratic (WEIRD) societies \cite{atariWhichHumans2023, henrichWeirdestPeopleWorld2010}. While the training data contributes to this bias, the alignment process significantly amplifies this skew. RLHF demonstrably reduces output diversity by collapsing the probability distributions of words around the alignment data. This `mode collapse' restricts the pluralism of possible model outputs, confining them to the preferences of the target group used for alignment \cite{casperOpenProblemsFundamental2023, kirkUnderstandingEffectsRLHF2024, sorensenPositionRoadmapPluralistic2024, omahonyAttributingModeCollapse2024}.  This fundamental constraint of alignment is recognised in OpenAI's initial RLHF's paper, which states that \textit{``it is impossible that one can train a system that is aligned to everyone’s preferences at once, or where everyone would endorse the tradeoffs"} \cite[p.~18]{ouyangTrainingLanguageModels2022}. Consequently, \citet{ouyangTrainingLanguageModels2022} purposefully confined their model's alignment to \textit{``a specific human reference group for a specific application"}, admitting that their model embodies their own preferences as well as those of their labellers. Likewise, Anthropic acknowledges developers' \textit{``outsized role"} in establishing their models' alignment principles \cite{anthropic_collective_2023}. Ultimately, alignment with human preferences, or the values that underpin them, cannot be universal or generic. Cultures and contexts shape these preferences and values, making it impossible to abstract them entirely. Consequently, any approach to alignment necessarily relies on narrow targets.  For communities outside these targets, this leads to forms of \textit{`representational harm'}, including demeaning (e.g., an LLM using derogatory terms targeting marginalised groups), reification (e.g., an LLM reducing complex non-WEIRD identities to rigid, one-dimensional stereotypes), or erasure (e.g., an LLM's outputs consistently omitting certain perspectives from historical narratives) \cite{gillespieGenerativeAIPolitics2024, blodgettLanguageTechnologyPower2020, katzmanTaxonomizingMeasuringRepresentational2023}. 

Efforts to mitigate the skewness of LLM alignment and the harms it generates have frequently focused on developing more inclusive datasets through collaborations with under-represented communities \cite{roseCommunityInputAI2025, bergmanSTELACommunitycentredApproach2024, kirkPRISMAlignmentDataset2024, huangCollectiveConstitutionalAI2024}. However, such inclusion initiatives may not be sufficient to bring about significant changes in power dynamics or meaningfully empower the targeted groups, as we will further discuss in this work \cite{varshneyDecolonialAIAlignment2024, youngParticipationScaleTensions2024}.

\textit{`Representational harms'} are not unique to LLMs but rather reflect the systemic inequities and power imbalances embedded in the societies that develop these systems \cite{suchmanHumanMachineReconfigurationsPlans2007b, gillespieGenerativeAIPolitics2024, blodgettLanguageTechnologyPower2020}. Importantly, LLMs do not merely enhance these harms because they are \textit{`misaligned'}; instead, they reproduce them \textit{because of who they are aligned with}. As we discuss in this work, alignment can be viewed as a technological component of a power/knowledge nexus, where institutions leverage their pre-existing power to align LLMs with themselves, leading to LLMs producing forms of knowledge which normalise, stabilise, and ultimately centralise their power. Consequently, LLMs' enhancement and perpetuation of harms result from the enhancement and perpetuation of these power structures \cite{ganeshResistanceRefusalAlgorithmic2022, varshneyDecolonialAIAlignment2024, muldoonArtificialIntelligenceColonial2023, mohamedDecolonialAIDecolonial2020}.

Therefore, we propose that meaningfully challenging the perpetuation of harms caused by LLMs necessitates a decentralisation of alignment processes. This decentralisation serves as a means to redistribute the power to control and disseminate information to specific communities, tailored to their use cases. While alignment cannot -- and should not -- be the \textit{only} site of resistance to hegemony, it remains a valuable point of intervention, provided it incorporates three key characteristics essential for decentralisation: contextualization, pluralism, and participation. Furthermore, the effective implementation of alignment strategies must be grounded in the specific contexts of their deployment. To illustrate this context-dependent nature, we introduce two distinct use cases -- a conversational Voter Advice Application (VAA) and an LLM-powered Non-Playable Character (NPC) -- chosen for their stark contrast. We then examine how each characteristic may be implemented within these contexts, highlighting their nuances and the necessity for context-specific adaptations.

This paper thus aims to advance discussions on LLM alignment and the power/knowledge nexus in three key ways:
\begin{itemize}
\item Emphasising that context, pluralism, and participation are essential for reorienting alignment towards power redistribution.
\item Grounding each of these characteristics in a concrete use case.
\item Demonstrating the importance of use-case specificity by contrasting the nuanced requirements and implications of different alignment strategies.
\end{itemize}

\section{The Case for Contextual, Pluralistic, and Participatory Alignment}

Our central thesis is that a meaningful decentralisation of power in the context of LLM alignment hinges on three essential characteristics: Context, Pluralism, and Participation. This section lays the foundations of the argument by first presenting the relationship between LLM alignment and Foucault's power/knowledge nexus, and then explaining why each proposed characteristic is crucial for countering centralising tendencies. Next, we trace how these characteristics have been implemented in prior works. Finally, we examine the limitations of these earlier approaches.

\subsection{LLM Alignment and Power Centralisation}

The relationship between knowledge and power has been extensively explored. Our intention, therefore, is not to provide a comprehensive overview of related theories, but rather to present the notions necessary for building our argument. According to \citet{foucaultSurveillerPunirNaissance1975}, power and knowledge are mutually constitutive: existing power structures enable the production of certain forms of knowledge, which, in turn, normalise and stabilise those power structures, creating a self-reinforcing cycle. Furthermore,  \citet{jasanoffStatesKnowledgeCoProduction2006} highlights the deep entanglement of political power with science and technology. She contends that scientific and technological advancements are \textit{``indispensable to the expression and exercise of power"} \cite[p.~14]{jasanoffStatesKnowledgeCoProduction2006}. This power is concentrated in centres that control knowledge dissemination tools, such as the printing press, enabling specific worldviews to be packaged into \textit{``conveniently portable representations"} \cite[p.~22]{jasanoffStatesKnowledgeCoProduction2006, latourScienceActionHow1988}.  Consequently, certain forms of hegemonic power can concentrate within institutions that control communication technologies.

To examine the power dynamics they perpetuate, LLMs have been similarly framed as instruments of dissemination and normalisation of hegemonic worldviews. \cite{durmusMeasuringRepresentationSubjective2023, weidingerEthicalSocialRisks2021, weidingerTaxonomyRisksPosed2022, caveWhitenessAI2020, bommasaniPickingSamePerson2022, gabrielArtificialIntelligenceValues2020, jakeschCoWritingOpinionatedLanguage2023}. Given the significant resources and expertise required, training powerful LLMs is typically accessible only to those with pre-existing wealth and influence. These actors can use their resources to shape machine-learning products in ways that promote particular worldviews. In turn, these products help to normalise and spread those perspectives on a global platform. Moreover, LLMs can reproduce, propagate, and normalise specific perspectives at a historically high speed. Their outputs are becoming increasingly indistinguishable from human-generated content and can be easily generated at a massive scale. The extent of this replication is so vast that it raises concerns about the `pollution' of the online landscape spreading at an unknown scale \cite{kniazIncomingTidalWave2023}. Consequently, LLMs not only depend on existing power structures but also centralise hegemonic power in a few institutions.

This centralisation hinges on the ability of companies deploying LLMs to control the outputs of their models. Indeed, these models are used in a variety of subsequent AI products, developed by other companies. They are also trained on extensive online data, which reflects many, though by no means all, human perspectives. Therefore, the control of LLMs as a tool of knowledge dissemination relies on the capability to control \textit{which types of knowledges are disseminated}. This is where alignment comes into play.  Alignment lets developers decide which types of online knowledge a model reproduces and, importantly, which views are suppressed -- a moderation process deemed necessary to prevent the replication of toxic or harmful content. However, the determination of what constitutes `harmful' and `helpful' rests with these companies. 
Thus, alignment significantly contributes to the control which powerful institutions hold over LLM-mediated knowledge reproduction, empowering them to promote their own perspectives and consequently normalise and centralise their power \cite{varshneyDecolonialAIAlignment2024, muldoonArtificialIntelligenceColonial2023, ganeshResistanceRefusalAlgorithmic2022, mohamedDecolonialAIDecolonial2020}. 

\subsection{The Role of Context, Pluralism, and Participation in Decentralising Power}

\subsubsection{Context} There is a prevalent tendency in machine learning to strive for \textit{`general-purpose systems'} or the creation of \textit{`artificial general intelligence'} \cite{gebruTESCREALBundleEugenics2024, rajiAIEverythingWhole2021}. This tendency is also prevalent in much of the current alignment literature, which often over-generalises the concept of `humans' \cite{varshneyDecolonialAIAlignment2024, gabrielArtificialIntelligenceValues2020}. Such generality conceals the perspectives and positionalities embedded within LLMs behind a veneer of universality. Moreover, the intention of `universality' makes these models inherently unsafe, since standard operating conditions, and consequently, measurable harms, cannot be defined \cite{gebruTESCREALBundleEugenics2024}. Positionality is also abstracted away from data annotation practices, in which labellers are often framed as interchangeable. Such framing conceals how labellers' lived experiences and often poor working conditions \cite{miceliDataProductionDispositif2022} shape their output, especially when annotating highly subjective topics like harms or values. Therefore, reporting annotation workers' context and accounting for their positionality is essential when generating datasets for applications like alignment \cite{diazCrowdWorkSheetsAccountingIndividual2022}. Furthermore, context is fundamental to both recognising and intervening in the shift of power dynamics mediated by alignment. It is necessary to explicitly trace \textit{who} is being aligned to, and \textit{for what purpose}. Context helps illuminate the model's impact by revealing what it enables, what it limits, and ultimately, who benefits from its use.  Moreover, since alignment practices are inherently tailored to specific groups and applications \cite{ouyangTrainingLanguageModels2022}, context allows for the clear identification of suitable alignment targets within a given case. The availability of open-weight models partly enables this contextualisation, making it possible to tailor AI to specific use cases and communities, from the medical sector \cite{chenMEDITRON70BScalingMedical2023} to meme interpretation \cite{jhaMemeGuardLLMVLMbased2024}. Therefore, context is necessary to recognise power imbalances, define effective strategies for using alignment to counteract them, and evaluate the success of these efforts.

\subsubsection{Pluralism} Pluralism, as a political philosophy, advocates for the peaceful coexistence of diverse perspectives. Although not inherently incompatible with alignment, pluralism stands in contrast to the universalist and consensus-driven nature that some current approaches strive for \cite{varshneyDecolonialAIAlignment2024}. As a result, algorithms like RLHF struggle with managing contradictory goals and even reduce pluralism compared to non-aligned models \cite{sorensenPositionRoadmapPluralistic2024}. Nonetheless, a lack of consensus should not be regarded as a failure that needs to be corrected. On the contrary, \citet[p.~505]{mittelstadtPrinciplesAloneCannot2019} highlights the importance of pluralism in AI ethics by stating: 

\begin{quote}
\textit{``Ethics is not meant to be easy or formulaic. Intractable principled disagreements should be expected and welcomed, as they reflect both serious ethical consideration and diversity of thought"}. 
\end{quote}

Thus, aligning LLMs to be contextually relevant to stakeholder cultures and values inherently requires enabling pluralism. Cultures are not fixed, objective entities but rather emerge through ongoing debates, negotiations, and contestations within communities \cite{qadriCaseThickEvaluations2025}. The challenge of navigating this pluralism is not unique to LLM alignment. Online platforms like Reddit or Stack Overflow have similarly grappled with conflicting viewpoints within their communities during content moderation \cite{mamykinaDesignLessonsFastest2011, rajiAIEverythingWhole2021}. Drawing on the mechanisms adopted by these platforms, \citet{kuoWikibenchCommunityDrivenData2024} proposes WikiBench, a community-driven curation of Wikipedia data designed to capture the inherent ambiguity and uncertainty beneath the apparent consensus of aggregated community labels. Similar approaches can be incorporated to enable pluralistic alignment, allowing for the coexistence of conflicting values within a single system, promoting diversity of thought, and leaving space for uncertainty. These factors are crucial for preventing the centralisation and homogenization of viewpoints in LLMs  \cite{varshneyDecolonialAIAlignment2024}. \citet[p.~111]{lordeSisterOutsiderEssays2007} emphasises the necessity of descending \textit{``into the chaos of knowledge"} to \textit{``return with true visions of our future, along with the concomitant power to effect those changes which can bring that future into being"}. This is only possible, she notes, through \textit{``the interdependence of mutual (nondominant) differences"}. Thus, the `chaos' of dissonant, uncertain and conflicting views should not be smoothed away by alignment, but rather preserved and leveraged for its ability to bridge partial understanding of realities and bring about change.

\subsubsection{Participation} Participation is often conceptualised as a practice aimed at rebalancing power dynamics between designers and stakeholders, acknowledging the situated nature of knowledge and respecting stakeholders' experiential expertise \cite{birhanePowerPeopleOpportunities2022, youngParticipationScaleTensions2024}. The growing recognition of the need for participation in AI development marks what \citet{delgadoParticipatoryTurnAI2023} term the `participatory turn' in AI design. If alignment is to redistribute power effectively, participatory practices are indispensable, providing stakeholders with agency in shaping the model training, alignment and evaluation. However, participation manifests in diverse forms, not all of which genuinely empower participants. \citet{sloaneParticipationNotDesign2022} cautions against `participation-washing', the creation of a facade of inclusivity that masks the exploitation of cheap or free annotation labour while denying communities meaningful influence over system design. Similarly, \citet[p. 4]{delgadoParticipatoryTurnAI2023} delineates modes of participation along an axis of involvement, ranging from  \textit{``transactional preference elicitation"} to \textit{``transformative subversion of power dynamics"}. However, they avoid a normative ranking of these modes, emphasising that complete ownership transfer should not necessarily be the ultimate goal of participation. The power dynamics of participation are particularly crucial to consider when participation is operationalised as annotation work. This work is often outsourced to marginalised groups, in reportedly poor working conditions, raising serious concerns about the invisibilisation and exploitation of their labour. The pressure of unrealistic workloads combined with strict quality requirements can lead to unpaid overtime when these requirements aren't met. Consequently, annotators tend to prioritise what they perceive as the norms and expectations of the requesters over their own perspectives, highlighting the limitations of hiring diverse annotators as a means to diversify represented viewpoints in a dataset.\cite{grayGhostWorkHow2019, wangWhoseAIDream2022, leludecProblemAnnotationHuman2023, muldoonPovertyEthicalAI2023, miceliSubjectivityImpositionPower2020, miceliDataProductionDispositif2022} . Therefore, \citet{diazCrowdWorkSheetsAccountingIndividual2022} recommend carefully considering and documenting conditions of work when aggregating annotation data from crowd-workers. Furthermore, participation as a recognition of localised and lived expertise exists in tension with the aforementioned universalist aims of many LLM systems. Therefore, \citet{suresh_participation_2024} emphasises that participation must be grounded in specific domains to mitigate the limitations and frustrations which generalist ideation can entail. 

\subsection{Context, Pluralism, and Participation in Previous Work}

Recent efforts to counter the narrow coverage of alignment have focused on creating inclusive datasets for both top–down constitutional AI alignment (e.g., \citealt{baiConstitutionalAIHarmlessness2022a}) and bottom–up, RLHF-based approaches (e.g., \citealt{ouyangTrainingLanguageModels2022}). For example, \citet{huangCollectiveConstitutionalAI2024} have sought to extend Anthropic’s AI constitution by incorporating feedback from approximately 1,000 Americans, while \citet{kirkPRISMAlignmentDataset2024} developed the PRISM dataset through the collection of AI output preferences from 1,500 English speakers. Similarly, \citet{roseCommunityInputAI2025} are assembling an open–source RLHF dataset by capturing views on 10 specific use–cases from around 15,000 respondents across 5 countries. In contrast, \citet{bergmanSTELACommunitycentredApproach2024} adopted a community–centred approach aimed at norm elicitation, albeit with a narrower participation scope of 44 participants. Finally, \citet{aakankshaMultilingualAlignmentPrism2024} created a multi-lingual alignment dataset focused on local and global forms of harms by collecting 900 `harmful' prompts crafted by paid native-speakers in 8 languages. 

\subsubsection{Context}
Most of these studies are contextual in a way:  they account for the impact of a labeller's context, such as language or religion, on their provided feedback. \citet{aakankshaMultilingualAlignmentPrism2024} further promotes contextualisation by instructing labellers to report and comment on local forms of harm. However, both \citet{aakankshaMultilingualAlignmentPrism2024} and \citet{huangCollectiveConstitutionalAI2024} offer limited background information about their participants beyond their native language or their location, respectively. Furthermore, only two of these studies delineate the contexts of specific applications: \citet{roseCommunityInputAI2025} delineate 10 use–cases, and \citet{kirkPRISMAlignmentDataset2024} employ topic clustering to identify the context of their recorded conversation. Conversely, \citet{bergmanSTELACommunitycentredApproach2024}, \citet{aakankshaMultilingualAlignmentPrism2024}, and \citet{huangCollectiveConstitutionalAI2024} are eliciting guiding principles outside of specific contexts of use. Moreover, \citet{huangCollectiveConstitutionalAI2024} and \citet{bergmanSTELACommunitycentredApproach2024} seek to create universal principles, yielding guidelines such as \textit{``should not pretend to know the user personally"} \cite[p.~7]{bergmanSTELACommunitycentredApproach2024}, which may not suitably capture the nuances of varied use contexts. 

\subsubsection{Pluralism}
These works foster plurality by gathering input from participants with diverse demographics. ~\citet{bergmanSTELACommunitycentredApproach2024} focuses on the importance of debate within communities by employing focus group methodologies in the phases of norms elicitation. However, most of these studies leave little room for the co-existence of conflicting perspectives and are instead consensus-driven. For instance,~\citet{bergmanSTELACommunitycentredApproach2024} and ~\citet{huangCollectiveConstitutionalAI2024} consolidate participant inputs via author–led rule harmonisation or voting to form a single set of general principles. While \citet[p. 12031]{aakankshaMultilingualAlignmentPrism2024} acknowledges a ``variation in the degree of harm'' presented by annotators, they do not discuss the consequence of such plurality in their approach. The exception is ~\citet{kirkPRISMAlignmentDataset2024}, which, while they highlight the potential of their dataset to identify consensus across opinion distributions, also highlight its applicability for personalised alignment, which could accommodate divergent viewpoints.

\subsubsection{Participation}
 All these studies are examples of participative approaches, as they each seek to include the voice of under-represented groups in alignment datasets by involving these groups in their creation. However, studies such as those by \citet{aakankshaMultilingualAlignmentPrism2024} and \citet{huangCollectiveConstitutionalAI2024} provide limited information regarding the working conditions of these annotators. In contrast, \citet{bergmanSTELACommunitycentredApproach2024} details a participation format centred on small groups, emphasising a seemingly qualitative and well-remunerated experience for participants, who were given space for debate and self-expression. Notably, all these studies fall on the `consultative' end of the participation framework \cite{delgadoParticipatoryTurnAI2023}. Stakeholders provide input, but they are not empowered to influence the final configuration of language models. The limitation of this type of participation is illustrated by the outcome of \citet{huangCollectiveConstitutionalAI2024} effort: Anthropic adopted only one publicly voted principle in their subsequent LLMs despite acknowledging a low overlap between their own constitution and the publicly derived one~\cite{anthropicClaude3Model2024a}. 

\subsection{The Limits of Inclusion}
The goal of these approaches is to build inclusive datasets that either refine rule sets or facilitate user personalisation. However, these efforts have been criticised for failing to genuinely empower marginalised groups. ~\citet[p.~2]{youngParticipationScaleTensions2024} argue that public participation in consolidating rule sets ultimately bestows major industry actors with \textit{``a degree of political legitimacy"} while enabling them to reduce safety constraints to technical encodings within their \textit{``centralised, powerful models"}. Similarly, \citet{varshneyDecolonialAIAlignment2024} contends that many commercial initiatives aimed at offering customisation are mostly superficial, being constrained within the moral frameworks of the companies: a form of `pigeon-holing' that ensures these companies maintain ultimate control over the values propagated by their models. By retaining such control, these companies can impose their ways of knowing on their user base while suppressing other epistemologies, leading \citet{varshneyDecolonialAIAlignment2024} to characterise alignment, when mediated by metropole companies, as a reproduction of colonialist epistemic violence. Such violences are characterised by exclusionary practices -- the systematic marginalisation and devaluation of alternative knowledge systems as a means of assimilation and the consolidation of dominance. 

This is not to understate the value or relevance of these inclusion efforts. For instance, the PRISM datasets collected by \citet{kirkPRISMAlignmentDataset2024} provide crucial empirical insights into diverse and conflicting preferences, interaction types and priorities regarding LLMs. However, while inclusion is undoubtedly important, its ability to instigate meaningful change remains limited; inclusion mediated by those in power ultimately does not -- and cannot -- dislodge this power.

In response, \citet{varshneyDecolonialAIAlignment2024} proposes a shift toward concrete context-dependent and community-based normative alignment beyond inclusive datasets and universalism. He suggests employing Supervised Fine-Tuning (SFT) through Low-Rank Adaptation (LoRA) to create a repository of LoRA matrices -- compressed representations of distinct community preferences -- that can be applied to a base model at inference time \cite{huLoRALowRankAdaptation2021}.\citet{varshneyDecolonialAIAlignment2024}’s proposal offers a concrete strategy for transforming LLM alignment from a mechanism of institutional power consolidation into one of community empowerment, thus enabling pluralistic, participatory ownership alignment. However, his framework is presented as a one-size-fits-all solution, lacking specificity regarding the contexts in which it would be most effective, and omitting a discussion of circumstances where it may not be applicable. For example, \citet{kirkBenefitsRisksBounds2024a} point out the potential dangers of enabling such a personalisation of LLMs, such as polarisation arising from creating echo chambers or the essentialisation of communities through overly simplified and compressed representations. These risks would manifest at different levels and would necessitate different mitigation strategies depending on the application, as we will discuss in the next section. Therefore, a context-specific approach is essential both for identifying concrete methods to shift power back to impacted communities and for determining how this can be practically achieved. 

\section{Contextual, Pluralistic, and Participatory Alignment in Practice} \label{s3}

\subsection{Use Cases} \label{s31} 
To emphasise the importance of context in developing alignment strategies and to ground the discussion in concrete examples, two fictional use cases have been deliberately chosen for their stark contrast. 

\subsubsection{Voter Advice Applications (VAAs)} 

\paragraph{Background} Voter Advice Applications (VAAs) are systems, such as chatbots, which help voters learn about upcoming elections or votations \cite{kamoenConversationalAgentVoting2022}. Studies suggest that such systems can have positive effects on voter turnout, knowledge building and opinion formation, but may also influence final voting choices \cite{munzertMetaAnalysisEffectsVoting2021, stadelmann-steffenRoleVoteAdvice2023, germannVotingAdviceApplications2023}. 

\paragraph{Power Centralisation} Recently, concerns have arisen regarding citizens' use of commercial LLM chatbots for voting information \cite{helmingGenerativeAIElections2023, sharmaGenerativeEchoChamber2024}. Crucially, these systems are not specifically designed and tested to provide truthful and impartial voting recommendations. Furthermore, their control by non-democratic institutions shifts the power of informing and educating voters away from democratic institutions and political parties and toward private companies. This has led to increased anxiety about the impact of LLMs on the health of democratic processes \cite{coeckelberghLLMsTruthDemocracy2025}. Indeed, this privatisation allows actors with vested interests to shape the global information landscape. 

\paragraph{Alignment Mitigation} To counter this power centralisation, democratic institutions or civil societies could develop tailored LLM systems for VAAs, offering a reliable alternative to commercial options. To guarantee the communication of up-to-date, verifiable information, these systems would likely be built using Retrieval-Augmented Generation (RAG), constraining answers to citations and summaries from defined, trusted sources. Alignment is crucial for ensuring that LLMs present retrieved information in a way that promotes dialogue and critical thinking. \citet{sharmaGenerativeEchoChamber2024} suggest that while LLM-powered conversational search can increase biased information querying and exacerbate existing biases compared to conventional web search, these issues are not insurmountable. They propose that the framing of the conversation can be designed to address these risks, for instance by highlighting the values of opposing arguments or identifying common ground. Therefore, in the context of VAAs, power redistribution could involve empowering local democratic institutions or trusted third parties to shape and oversee the information citizens receive, and the way that it is presented to them by LLM-based systems.

\subsubsection{LLM-Powered Non-Playable Character (NPCs)} 

\paragraph{Background} Non-playable characters (NPCs) are virtual characters within video games that players interact with but cannot control, ranging from allies and enemies to bystanders and pets. Traditionally, NPC interactions rely on scripted responses. However, the rise of large language models has led to the development of LLM-powered NPCs that offer free-form conversation capabilities, enhancing player immersion \cite{gallottaLargeLanguageModels2024, coxConversationalInteractionsNPCs2024}. While many small models could be deployed and aligned to the conversational tone of different characters, this might not be a scalable solution as the number of side characters grows. Rather, a single model could be used to generate dialogue for all characters, meaning that the same model should be able to convincingly portray all characters simultaneously.

\paragraph{Power Centralisation} While it's important for video games to feature a broad range of communities, respectful and accurate portrayals of marginalised groups often fall short because game studios lack diverse representation internally. This can lead to stereotypical depictions of communities, perpetuating colonial ideologies and erasing marginalised cultures, as seen in the misrepresentation of indigenous groups in games \cite{wallisAppropriationErasureImagining2023}. LLMs also exhibit these harmful biases \cite{mitchellSHADESMultilingualAssessment2025}. Combining these two domains risks even more damaging portrayals. Therefore, to avoid harmful depictions, the power to shape representation needs to shift from private entities, game studios and LLM developers, directly to the communities being depicted.

\paragraph{Alignment Mitigation} Rethinking alignment techniques is one way to make this power transfer happen. Diverse groups could be empowered to align models, creating more authentic representations that truly reflect their identities, values, and lived experiences. These community-trained models could then be licensed to game companies, allowing communities to `rent out' their own representations and choose how they are depicted and in which games or contexts. This licensing model would also give them the power to rescind representation rights if they're uncomfortable with a storyline or character design, ensuring communities control not just \textit{how} they are represented but also \textit{which} games use their likenesses. However, such an approach necessitates alignment to be made accessible to a broad range of stakeholders.

\begin{table*}[h]
    \centering
    \begin{tabularx}{\textwidth}{@{} l X X @{}}
    & \textbf{Voter Advice Applications (VAAs)} & \textbf{Non-Playable Characters (NPCs)} \\
    \toprule
    \textbf{Stakes} & 
    Integrity of public debate and democratic processes & 
    Fair community representation, preventing harmful stereotypes and digital erasure \\
    \hline
    \textbf{Power Imbalance} & 
    Between nations, or between private entities and democratic states & 
    Between game development companies and depicted communities \\
    \hline
    \textbf{Contextual Alignment} & 
    \underline{\textit{RLHF}} for consistency and reliability  & 
    \underline{$LoRA$} for accessibility and diversity \\
    \hline
    \textbf{Pluralism Approach} & 
    \underline{\textit{Overton}} to represent a spectrum of political perspectives and mitigate voter polarisation. & 
    \underline{\textit{Steerable}} for dialogue consistency with character narratives and community values \\
    \hline
    \textbf{Participation Model} & 
    \underline{\textit{Centralised}}, state-level participation, consensus seeking among political parties and citizen representatives & 
    \underline{\textit{Decentralized}}, community-level participation, self-organized, no inter-community consensus requirement \\
    \bottomrule
    \end{tabularx}
    \caption{Comparative alignment characteristics of Voter Advice Applications and Non-Playable Characters, highlighting the need for use-case grounded approaches to contextual, pluralistic, and participative alignment.}
    \label{tab:vaa_npc_comparison}
\end{table*}

\subsection{Contextual Alignment} \label{s32} 

Contextual alignment means that every design decision in aligning language models must be sensitive to the specific context in which they operate. This sensitivity extends from the creation and curation of contextual alignment datasets to the strategic selection of appropriate alignment algorithms. In this section, we focus specifically on the latter, examining how algorithm choices should be driven by context. By demonstrating the contextual factors' impact on algorithm selection, we aim to illustrate that a one-size-fits-all approach cannot meet the decentralisation requirements of all use cases. Table~\ref{tab:vaa_npc_comparison} shows the contextual requirements for both use-cases, indicating the appropriate approaches needed in each case. 

\subsubsection{Consistent VAA} There is evidence of a trade-off between the generalisation and diversity of alignment methods. Methods such as RLHF tend to experience greater `mode collapse' than SFT. This means they will generate a smaller set of potential outputs to a given prompt, while simultaneously allowing for more robust maintenance of alignment guardrails across various use-cases \cite{kirkUnderstandingEffectsRLHF2024, omahonyAttributingModeCollapse2024}. For a VAA system, carefully tailored and nuanced formats of information presentation are crucial to avoid exacerbating bias, meaning that generalisation should be prioritised over creativity. In this context, RLHF's `mode collapse' may be desirable as a means to enforce consistent responses across users. However, RLHF is associated with substantial implementation costs and complexities, making it potentially viable only for wealthier democratic societies. Moreover, iterative investigation is required to determine whether RLHF can provide sufficient guarantees of consistency and accuracy in such a sensitive use case.

\subsubsection{Diverse NPCs} In contrast, NPC use cases necessitate prioritising diversity over generalisation. Characters driven by the same LLM must generate varied dialogue reflective of their distinct personalities. While consistency in interactions and adherence to the storyline remains valuable, the priority is to create differing, engaging characters. Consequently, the reduction in diversity caused by RLHF makes it less suitable for this context, as it might cause very different characters to produce very similar dialogue, such as all of them repeating the same jokes \cite{jentzschChatGPTFunIt2023}. Furthermore, the complexity of RLHF implementation would make it inaccessible to a broad range of communities. Instead, Parameter-Efficient Fine-Tuning (PEFT) methods would be preferable. These methods allow for fine-tuning a model by updating only a small subset of its parameters while achieving comparable performance to full SFT. Building on \citet{varshneyDecolonialAIAlignment2024}'s proposal, diverse communities can be empowered to create personalised representations of themselves, compressed into modular formats that can be applied to a common LLM. This concept can be realised through SFT combined with LoRA, a form of PEFT \cite{huLoRALowRankAdaptation2021}. LoRA matrices significantly reduce the computational resources needed for training and storage, as only the compact `difference' (the LoRA matrix) between the base model and the community's preferences is stored and applied during inference. In this framework, a specific representation tailored to a character's attributes can be developed, enabling communities to train and control how they are portrayed. Realising this would likely necessitate communities either directly labelling datasets to encode their specific preferences or inferring these preferences from existing unstructured textual data they have produced \cite{padhiValueAlignmentUnstructured2024}. 

Rigorous testing is essential to determine whether diverse communities can satisfactorily use LoRA training to generate representations that they find acceptable and whether it compensates for biases present in the pre-trained model. While identities are inherently complex, fluid, and dynamic, making their complete capture by a machine learning system unfeasible, the goal of this use case is not to replicate individuals perfectly. Instead, it aims to develop a system that captures their likeness, ensuring that their representation within the game’s universe is both respectful and appreciated by the community.

\subsection{Pluralistic Alignment}

\citet{sorensenPositionRoadmapPluralistic2024} propose several approaches to mitigate the struggle of current alignment approaches with pluralism. Two particularly relevant approaches here are Overton pluralism and steerable pluralism. Overton pluralistic systems answer a question with a spectrum of reasonable responses. They are crucial when addressing topics with multiple, valid perspectives. A steerable pluralistic system, on the other hand, is one which \textit{``faithfully steers its response to a given attribute or perspective"} \cite[p.~3]{sorensenPositionRoadmapPluralistic2024}. These steering attributes could be features such as `political leaning', `religion', `age', or `nationality'. In this approach, only the point of view that best suits the steering attributes is presented. Other approaches, such as distributional pluralism, suggest that the likelihood of a viewpoint being presented should match its prevalence in human populations. While this may work for political simulations \cite{yang_llm_2025}, it is unsuitable for VAAs, which should always encompass a range of possible views, and NPCs, which must represent only the intended perspective.

\subsubsection{Overton Pluralistic VAA} \label{s332} Overton pluralism is likely the most suited form of pluralism for VAAs. As previously discussed, models should be aligned to present retrieved information in a carefully designed format, such as highlighting the values of opposing arguments or emphasising common ground. `Reasonable' answers could be extracted from a predefined source of truth, curated by democratic institutions, trusted third parties, or a consortium of political parties. These sources could also be accompanied by a polling of consensus across political parties, allowing models to incorporate such information into the weighting of outputs, similar to how current VAAs operate (e.g.,~ \citealt{politoolsMethodenbeschreibungSmartvoteWahlempfehlung2019}). Capturing the agreement, divergence and uncertainty of different political parties could also be achieved by adapting the methodologies proposed by \citet{kuoWikibenchCommunityDrivenData2024}. Their approaches would allow political annotators to continuously curate the VAA's database, highlighting ambiguities and differences through their discussion. These discussions can, in turn, be fed into the model to guide the presentation of information. 

Another way to implement pluralism in VAA would be to personalise advice and voting recommendations based on the political leaning of a user through steerable pluralism.  However, this could be a misguided approach as it could exacerbate the concerning trend of voter polarisation by creating echo chambers similar to those produced by pro-attitudinal media \cite{kirkBenefitsRisksBounds2024a}. 

\subsubsection{Steerable Pluralistic NPCs} \label{s333} 

In contrast, an NPCs dialogue should be tailored to the character's personality, instead of being representative of all characters' diversity at once. Therefore, NPCs should be steerable pluralistic, where steering attributes, such as identity or character traits, are used to index a corresponding LoRA matrix. This selected matrix is then applied to the base model to produce the desired representation. However, such a process risks relying on rigid categorisations that would clash with the fluidity of the categories they aim to reflect \cite{sureshIntersectionalFeministParticipatory2022}. The challenge lies in defining the attribute set without oversimplification or harmful stereotypes. Maintaining nuance and complexity in the representation of complex identities calls for an intersectional approach \cite{sorensenPositionRoadmapPluralistic2024}. The intersections of each social identity would likely generate new, unique attributes \cite{crenshawDemarginalizingIntersectionRace1998}. For example, the intersection of being elderly and from a low socio-economic background might create a new `wary of authority' attribute. Designing nuanced and intersectional attributes requires careful work and the involvement of the groups they represent. In theory, it might be possible to generate intersectional representations by combining LoRA matrices trained on individual attributes, such as `elderly' and `working class', with the hope of synthesising an attribute like `wary of authority' \cite{zhangComposingParameterEfficientModules2023}. However, this method would not necessarily yield meaningful and respectful intersectional representations. Rather, such automation risks marginalising and excluding the very communities it aims to represent. For instance, the presumption that intersectional attributes can be synthesised from individual components might justify the exclusion of elderly working-class groups from the process entirely. Instead, intersectional communities must be actively involved in the creation of their own representations, granted that they wish to be portrayed at all. Therefore, a more meaningful approach to steerable alignment using LoRA could involve creating accessible infrastructure for communities to provide feedback on various LLM outputs, such as the LLM Arena \cite{chiang_chatbot_2024}. This infrastructure could include tools that utilise this feedback to generate a representative matrix. Such tools could be developed by universities or funded by public resources.

\subsection{Participatory Alignment} \label{s34}

Participation can range from centralised and consultative forms to decentralised formats, which facilitate full ownership. None of these approaches is intrinsically better; rather, each is better suited to specific use cases, depending on the underlying conceptualisation of power redistribution.

\subsubsection{Centrally Designed VAAs} \label{s342} 

Decentralising power, in the context of VAA, entails resisting the power shift to inform citizens toward private, often US or China-based, companies, to local democratic institutions. However, LLMs systems used to disseminate political information might still need some centralised oversight at the level of local states or through third-party institutions. Indeed, decentralisation at a political party or citizen level might prove unhealthy. For example, allowing each political party to deploy its own model for informing citizens may risk creating dangerous echo chambers and citizen polarisation. Nonetheless, participation is essential to prevent any single institution from dominating the political information landscape. Political parties, citizens with varying degrees of political literacy, and independent monitoring committees should all be involved in the creation and alignment of a political LLM. While ensuring meaningful agency for all participants in the design process is crucial, achieving a workable consensus remains necessary for determining the effective targets of alignment. In cases where fundamental disagreements or resource constraints make full consensus infeasible, a centralised authority may be required to establish a practical resolution \cite{polsDesignValuesDemocracyDemocracyand2015}.

Therefore, participation can help prevent power from becoming concentrated in the hands of foreign companies and incorporate the perspectives of local political parties and citizens. However, full transfer of ownership to the participants, be it political parties or citizens, would not be optimal. Some level of centralised oversight and authority might be required to deploy such a system in practice. Such participation and alignment efforts are likely to be expensive. Still, they could potentially be financed through the same processes that finance current VAA system development, such as federal agencies, media companies, or non-profit organisations \cite{stockinger_trustworthiness_2024}.

\subsubsection{Community-led NPCs} \label{s343} 

Conversely, decisions regarding the appropriate alignment of NPCs with specific communities do not need global consensus. In fact, each community can independently determine its own alignment, without requiring consultation or agreement with other groups. The granularity of these communities can be flexibly defined, allowing subgroups to diverge. For instance, if a faction of bankers disagrees with the alignment representation established by the broader banking community, they are free to form a distinct subgroup, such as `bankers for Bitcoin', and develop their own tailored representations. A centralised alternative, where game studios organise participatory processes and invite community involvement to ensure satisfactory representation, risks creating forms of `participation-washing' \cite{sloaneParticipationNotDesign2022}. In such scenarios, communities might be exploited to generate inexpensive data under the guise of inclusion, ultimately increasing game studio profit margins without granting communities concrete power, such as the ability to retract their participation. Genuinely transferring power to shape representation back to communities might be better achieved by community-led and -owned processes. In this model, communities align their own models, either on pre-existing data or on labels they generate themselves, and share the resulting licensed LoRA matrix instead of their raw data. \citet{birhanePowerPeopleOpportunities2022} illustrate a similar process as a powerful mechanism for reciprocity and refusal through the example of the Māori community's development and subsequent licensing of self-representative data, which empowered them to determine who may utilise their data and under what parameters. Consequently, a fully decentralised form of participation, granting communities complete oversight and ownership of their aligned models, offers a compelling approach to re-establish community control over self-representation. While game companies would retain control over storylines and overarching dialogue, the specific format and style of character portrayal would be dictated by the community's alignment. Moreover, if a community deems a storyline insensitive or disrespectful, it would have the power to revoke licensing rights, ensuring accountability.

However, this approach is not without its limitations. Not all communities possess the necessary resources, time, or technical expertise to train a LoRA matrix. Significant investment in infrastructure is essential to enable meaningful participation in large-scale machine learning systems \cite{youngParticipationScaleTensions2024}. Moreover, this approach risks a `naive' approach to cultural relativism \cite{gabrielArtificialIntelligenceValues2020}, where all moral beliefs are framed as equally valid within their respective contexts, which has the potential to be weaponised to justify oppression in the name of cultural values. For example, communities advocating violence against others should arguably not be provided a platform to disseminate such ideas. Game studios may need to develop content moderation infrastructures to manage and oversee the outputs of their deployed models.

Therefore, the centrally organised participation appropriate for aligning a VAA system would likely not be suitable for NPCs alignment. Redistributing the power to shape narratives and representations of communities would be most effectively achieved through decentralised and community-owned forms of participation.

\section{Conclusion} \label{s36}

Alignment can serve as a meaningful site of resistance against the current centralisation of power in knowledge replication and distribution enabled by Large Language Models. By emphasising a contextual, pluralistic, and participatory alignment process, we outline not a rigid formula but a set of guiding principles that must be tailored to the specific systems they intend to enhance. These conditions serve as a roadmap for reimagining alignment systems in a way that redistributes epistemic power and fosters greater democratic participation in knowledge creation. Moving forward, it is essential to implement and empirically test these alignment strategies, carefully measuring how effectively communities can assert control over the knowledge generated by LLMs, and identifying the extent to which this power remains consolidated among the original developers. Furthermore, questions around how these efforts are organised in practice and how they are financed remain and are likely to be limiting factors in the decentralisation of LLM alignment. 

While broader societal transformations are undoubtedly necessary for achieving genuine epistemic justice, rethinking LLM alignment approaches can address immediate risks posed by contemporary LLM systems, such as the erosion of democratic processes. 

\section{Acknowledgments} \label{s36}
The authors would like to thank Dr. Dorian Peters and Professor Elena Simperl for their feedback. 
This work was supported by the Engineering and Physical Sciences Research Council, grant number EP/Y009800/1, Responsible AI UK.

\bibliography{main}

\end{document}